\documentstyle[12pt]{article}
\begin{document}
\baselineskip=24pt
\title{Horizon Mass Theorem\\
\vspace{1in} }
\author{Yuan K. Ha \\
Department of Physics, Temple University \\
Philadelphia, Pennsylvania 19122 U.S.A. \\
yuanha@temple.edu\\
May 1, 2005\\
\vspace{1.9 in} }
\date{Honorable Mention\\
Gravity Research Foundation 2005\\}
\maketitle
\newpage
\begin{abstract}
{\bf A new theorem for black holes is found. It is called the
horizon mass theorem. The horizon mass is the mass which cannot
escape from the horizon of a black hole. For all black holes:
neutral, charged or rotating, the horizon mass is always twice the
irreducible mass observed at infinity. Previous theorems on black
holes are: 1. the singularity theorem, 2. the area theorem, 3. the
uniqueness theorem, 4. the positive energy theorem. The horizon
mass theorem is possibly the last general theorem for classical
black holes. It is crucial for understanding Hawking radiation and
for investigating processes occurring near the horizon.\\

}\\
\end{abstract}

\newpage
A black hole is a system which has the maximum possible gravitational potential energy. The mass of a
black hole depends actually on where the observer is. In the usual case where the observer is
at infinity, the mass observed is the asymptotic mass and this is the mass which appears in
the Schwarzchild solution. As the gravitational field of a black hole extends to infinity,
its gravitational energy extends similarly and contributes to its mass. Since gravitational
energy is always negative, the closer one gets to a black hole the less gravitational energy
one sees. As a result, the mass of black hole increases as one gets near the horizon. An exact
energy expression of a Schwarzchild black hole in
the quasilocal energy approach [1], the teleparallel equivalent formulation of general relativity [2],
and the gravitational redshift approach [3] has been found.
The total energy contained within a radius at coordinate $r$ is given by
\begin{equation}
E(r) = \frac{rc^{4}}{G} \left[ 1 - \sqrt{ 1 - \frac{2GM}{rc^{2} }}  \right],
\end{equation}
where $M$ is the mass of the black hole observed at infinity, $c$ is the speed of light and
$G$ is the gravitational constant. At the horizon, the mass of the black hole is found to be
twice its asymptotic mass. The concept of a horizon mass has been introduced; it is the mass
which cannot escape from the horizon of a black hole. The black hole energy formula is potentially
important in understanding astrophysical processes and black hole collisions.\\

In this paper, we would like to investigate first the
gravitational energy outside an electrically charged black hole
and derive an energy expression similar to the Schwarzchild case.
Such a Reissner-Nordstr\"{o}m black hole has a mass $M$ and
electric charge $Q$, with a radius $R = r_{+}$ according to a
distant observer stationed at infinity. The mass $M$ includes
electrostatic energy and hence it has a greater value than the
corresponding mass when the black hole is neutral. The total
energy for this charged black hole including electrostatic energy
contained within a radius at coordinate $r$ is derived in a
physical way by considering the escape of a photon outside the
surface of the black hole. The horizon mass of the charged black
hole reveals a surprising connection with the asymptotic mass of
the black hole when it becomes neutral. This result and a
corresponding one on the rotating black hole
lead inevitably to a general theorem on the horizon mass for all black hole cases.\\

The Reissner-Nordstr\"{o}m solution to Einstein's equation is given in coordinates
$(t, r, \theta, \phi)$ by the metric [4,5]
\begin{equation}
ds^{2} = A(r) c^{2}dt^{2} - A^{-1}(r) dr^{2} - r^{2}d{\theta}^{2} - r^{2}sin^{2}{\theta}d{\phi}^{2},
\end{equation}
where
\begin{equation}
A(r) = \left( 1 - \frac{2GM}{rc^{2} }+ \frac{GQ^{2}}{r^{2}c^{4}} \right).\\
\end{equation}
in cgs system. The metric describes a spherically symmetric and static spacetime with radial electric fields
and gravitational fields. A Reissner-Nordstr\"{o}m black hole is formed
when a photon of a given energy emitted just outside the horizon will have zero energy as it reaches infinity.
If a photon is emitted at coordinate $r$ with energy $\varepsilon_{r}$ and subsequently observed at infinity,
its energy is given by
\begin{equation}
\varepsilon_{\infty} = \varepsilon_{r} \sqrt{ 1 - \frac{2GM}{rc^{2}} + \frac{GQ^{2}}{r^{2}c^{4}} },
\end{equation}
This represents the gravitational redshift of the photon including the effect of electrostatic energy contribution.
Although the term containing the charge $Q$ in the square root has an opposite sign to the term containing
the mass $M$, the gravitational redshift is still stronger in this situation than in the neutral black hole
case since $M$ here has a greater value.
The difference between the energies of the photon at the two locations
is therefore
\begin{equation}
\varepsilon_{r} - \varepsilon_{\infty} =
                  \varepsilon_{r}\left[ 1 - \sqrt{ 1 - \frac{2GM}{rc^{2}} + \frac{GQ^{2}}{r^{2}c^{4}} } \right].
\end{equation}
As in the Schwarzchild case, the change in the photon's energy is a measure of the change of the
gravitational potential energy of the charged black hole itself.
To describe the complete behavior of the energy of the electrically charged black hole,
we consider a function $f(r)$ interpolating between the horizon of the black hole and infinity so that
the energy of the charged black hole also becomes a function of the coordinate $r$.
This energy expression gives the {\em total
energy} of the charged black hole contained in a spherical volume from the origin up to the coordinate
$r$ and is given by
\begin{equation}
E(r) = f(r)\left[ 1 - \sqrt{ 1 - \frac{2GM}{rc^{2}} + \frac{GQ^{2}}{r^{2}c^{4}} }\right].
\end{equation}
To determine the function $f(r)$, we introduce the following criteria:\
\begin{enumerate}
\item The total energy $E(r)$ is always positive. Thus $f(r)$ must be
      a positive function between $r_{+}$ and $\infty$.
\item The total energy $E(r)$ decreases smoothly between $r_{+}$ and $\infty$.
      This follows from the weakening of the gravitational redshifts of photons emitted at increasing
      large distances and subsequently observed at infinity. Thus the derivative $dE/dr$ is always negative.
\item At large distances, the total energy $E(r)$ approaches an asymptotic value.
      Thus $dE/dr \simeq 0$ at very large distances.
\end{enumerate}
Taking the derivative $dE/dr$ in Eq.(6) and subjecting it to the above conditions, while observing that
at very large distances, all $1/r^{2}$ terms and those of higher inverse powers are insignificant compared with the
$1/r$ terms, we find at large distances an equation for $f(r)$,
\begin{equation}
   \frac{df(r)}{dr} = \frac{1}{r} f(r) .
\end{equation}
The solution is found to be $f(r)$ = constant$\times r$. Note that the horizon radius is a natural cutoff
to all small distance behavior for the distant observer.\\

To determine the constant, we notice at large distances, the square root in Eq.(6) expands as
$1 - GM/rc^{2} + O(1/r^{2})$ , the energy of the charged black hole
should approach the asymptotic value $Mc^{2}$ as seen by the distant observer. Thus,\\
\begin{equation}
  E(r) \simeq f(r) \left( {\frac{GM}{rc^{2}}} \right) \rightarrow Mc^{2},
       \hspace{.2in} r \rightarrow \infty,
\end{equation}
giving\\
\begin{equation}
f(r) = \frac{rc^{4}}{G}.\\
\end{equation}
The overall energy expression for the charged black hole is therefore
\begin{equation}
E(r) = \frac{rc^{4}}{G} \left[ 1 - \sqrt{ 1 - \frac{2GM}{rc^{2}} + \frac{GQ^{2}}{r^{2}c^{4}} }\right].\\
\end{equation}

\vspace{.2in}
To evaluate the energy within the horizon of the charged black hole, we may set $r = r_{+}$ in Eq.(10).
The result is simple because the square root factor vanishes identically at the horizon.
For the Reissner-Nordstr\"{o}m black hole, the horizon radius is determined by the condition $A(r) = 0$ in Eq.(3).
This is equivalent to solving a quadratic equation involving $r$, with the two roots being $r = r_{+}$ and $r = r_{-}$.
The horizon radius is given by
\begin{equation}
r_{+} = \frac{GM}{c^{2}} + \frac{GM}{c^{2}} \sqrt{1 - \frac{Q^{2}}{GM^{2}} }.
\end{equation}
The horizon energy calculated from Eq.(10) is therefore
\begin{equation}
E(r = r_{+}) = \frac{r_{+}c^{4}}{G} ,
\end{equation}
or,
\begin{equation}
E(r_{+})  = Mc^{2} + Mc^{2} \sqrt{1 - \frac{Q^{2}}{GM^{2}} }.
\end{equation}
The horizon radius and the horizon energy reduce to their respective Schwarzchild black hole expressions
when $Q = 0$, with a corresponding decrease in the value of $M$.\\

To evaluate the energy exterior to the charged black hole, we may subtract the horizon energy from
the asymptotic mass, since energy has an additive property in a static spacetime. The result is
\begin{equation}
E(\infty) - E(r_{+}) =  - Mc^{2} \sqrt{1 - \frac{Q^{2}}{GM^{2}} }.
\end{equation}
This represents a net combination of negative gravitational energy
and positive electrostatic energy outside the black hole. The
gravitational energy in general dominates. A special case occurs
when the black hole has a maximal charge allowed by a
gravitational system, given by the condition $Q^{2} = GM^{2}$. The
electrostatic energy then is seen to cancel the gravitational
energy, and the energy outside such a black hole is identically
zero. When the horizon mass is the same as the asymptotic mass,
the extreme charged black hole cannot emit gravitational waves or
particles in the form of Hawking radiation [6]. As soon as it
emits a neutral particle, the black hole will have a smaller mass,
thus violating the condition $GM^{2} \geq Q^{2}$. This is not
allowed. If the black hole emits a particle with a charge and a mass
so that the condition $GM^{2} \geq Q^{2}$ is satisfied, an observer
stationed near the horizon and an observer stationed at infinity
both will reckon that the black hole loses the same mass
corresponding to the mass of the particle, however there is no
source of energy for the particle to escape to infinity if there
is no external agent or extra energy supply from the black hole.
Again this process is not possible. The extreme charged black hole
is therefore absolutely stable against emission. The effect becomes
important when quantum gravity is considered.\\

The irreducible mass of a charged black hole is the final mass of the black hole when its charge is
neutralized by allowing oppositely charged particles to be lowered slowly to the black hole, thereby
extracting energy from the black hole. The irreducible mass $M_{irr}$ does not include electrostatic energy.
It is always smaller than the initial mass of the charged black hole and it is defined by [7]
\begin{equation}
M = M_{irr} + \frac{Q^{2}}{4GM_{irr}} ,
\end{equation}
or equivalently,
\begin{equation}
M_{irr} = \frac{M}{2} + \frac{M}{2} \sqrt{1 - \frac{Q^{2}}{GM^{2}} },
\end{equation}
which implies the area non-decrease theorem for black holes [8].
The irreducible mass is an asymptotic Schwarzchild mass seen by a distant observer. For a Schwarzchild
black hole, recall the mass contained in the horizon is always equal to twice its asymptotic mass.
The negative gravitational energy exterior to the Schwarzchild black hole is as great as its asymptotic mass.\\

The horizon mass of a charged black hole is the mass contained
within the horizon as seen by a stationary observer just outside
the horizon. It is the mass which cannot escape from the horizon
of the charged black hole. Dividing the horizon energy by $c^{2}$
in Eq.(13), we obtain the horizon mass
\begin{equation}
M(r_{+}) = 2 M_{irr} .
\end{equation}
The horizon mass of the charged black hole is found to be simply
twice its irreducible mass. Note that it is {\em not} twice the
charged black hole mass $M$ which includes electrostatic energy.
This is a remarkable result. The horizon mass of the charged black
hole depends only on the energy of the black hole when it is
neutral! A logical explanation is that electric charges reside
only on the surface of the black hole as seen by a distant
observer. A charged black hole therefore behaves like a perfect
electric conductor. This is quite natural considering the
formation of a charged black hole by watching a charged mass
falling into a Schwarzchild black hole. To the distant observer,
the charged mass will never cross the horizon and appear to stay
at the surface of the black hole. The horizon mass in fact remains
constant as the
charged black hole becomes a neutral black hole in the neutralization process.\\

An analogous situation exists for the rotating black hole. The
irreducible mass of a rotating black hole is the final mass of the
black hole when its rotational energy is completely extracted by
interacting with external particles in the ergosphere [9,10]. The
irreducible mass is always less than the initial mass which
includes rotational energy. The rotating black hole then becomes a
Schwarzchild black hole. Analysis of the horizon mass of a
rotating black hole in the quasilocal energy approach [11] and in
the teleparallel equivalent formulation of general relativity [12]
reveals that it is extremely close to twice the irreducible mass.
This implies that there is very little rotational energy inside
the black hole. The result has to be obtained in an approximation
due to the complexity of the evaluation procedure and because the
Kerr metric has only axial symmetry rather than spherical symmetry
[13]. Again, the horizon mass of the rotating black hole appears
to depend only on the energy of the black hole when it is not
rotating! This suggests that the rotational energy resides
completely outside the black hole according to a distant observer
and that the horizon mass remains constant during the energy
extraction process. It explains why the horizon of a rotating
black hole has a constant radial coordinate independent of angle
whereas the outer boundary of the ergosphere shows a bulging which
depends on the angular momentum of the black hole. In fact, the
angular velocity inside the black hole cannot be defined in the
conventional way because spacetime is severely distorted there and
it is meaningless to visualize any internal rotation. It is a
paradox that the rotating black hole behaves as a perfect rigid
body with the same angular velocity everywhere on the horizon to an outside observer.\\

From the Kerr metric, it is further known that the area of a
rotating black hole at the event horizon when expressed in terms
of its irreducible mass $M_{irr}$ is [14]
\begin{equation}
A = \frac{16 \pi G^{2}M_{irr}^{2}}{c^{4}}.
\end{equation}
However, a locally nonrotating observer, who is seen to be
comoving with the black hole at the event horizon by a distant
observer, believes there is a Schwarzchild black hole. Such a
Schwarzchild black hole with the same surface area as that in
Eq.(18) has an asymptotic mass $M_{irr}$. Once again, since the
horizon mass of a Schwarzchild black hole is always twice its
asymptotic mass, therefore the horizon mass of the `rotating black
hole' is simply $2M_{irr}$. The result is valid for all allowed
angular momenta of what the distant observer calls the rotating black hole.\\

We may finally state a general theorem about black holes, whether they are neutral, charged, or rotating:\\

{\it For all black holes, the horizon mass is always equal to
twice the irreducible mass observed at infinity.}\\

The conclusion is striking. The electrostatic energy and the
rotational energy of a general black hole are all external
quantities. They are absent inside the black hole. The horizon
mass theorem stated above is crucial for understanding the
occurrence of Hawking radiation. No black hole radiation
whatsoever is possible if the horizon mass is the same as the
asymptotic mass. Without
black hole radiation, the second law of thermodynamics is lost.\\


\newpage
\begin{thebibliography}{99}
\bibitem{1} J.D. Brown and J.W. York, Jr., {\em Phys. Rev.} D {\bf47}, 1407 (1993).
\bibitem{2} J.W. Maluf, {\em J. Math. Phys.}  {\bf36}, 4242 (1995).
\bibitem{3} Y.K. Ha, {\em Gen. Rel. Grav.} {\bf35}, 2045 (2003).
\bibitem{4} H. Reissner, {\em Ann. Physik} {\bf50}, 106 (1916).
\bibitem{5} G. Nordstr\"{o}m, {\em Proc. Kon. Ned. Akad. Wet.} {\bf20}, 1238 (1918).
\bibitem{6} S.W. Hawking, {\em Commun. Math. Phys.} {\bf43}, 199 (1975).
\bibitem{7} D. Christodoulou and R. Ruffini, {\em Phys. Rev.} D {\bf4}, 3552 (1971).
\bibitem{8} S.W. Hawking, {\em Commun. Math. Phys.} {\bf25}, 159 (1972).
\bibitem{9} R. Penrose, {\em Rev. del Nuovo Cimento} {\bf1}, 252 (1969).
\bibitem{10} D. Christodoulou, {\em Phys. Rev. Lett.} {\bf25}, 1596 (1970).
\bibitem{11} E.A. Martinez, {\em Phys. Rev.} D {\bf50}, 4920 (1994).
\bibitem{12} J.W. Maluf, E.F. Martins and A. Kneip, {\em J. Math. Phys.} {\bf37}, 6302 (1996).
\bibitem{13} O.B. Zaslavskii, {\em Class. Quantum Grav.} {\bf12}, L 63 (1995).
\bibitem{14} R.M. Wald, {\em General Relativity}, (The University of Chicago Press, 1984).
\end{thebibliography}
\end{document}